\documentclass[a4paper,11pt]{article}
\usepackage{amssymb,amsmath}
\usepackage{cite,citesort,graphicx}
\newcommand{\gref}[1]{(\ref{#1})}

\newcommand{\bra}[1]{\left\langle #1 \right|}
\newcommand{\ket}[1]{\left| #1 \right\rangle}
\newcommand{\bk}[2]{\left\langle #1 | #2 \right\rangle}

\newcommand{\de}[2]{\frac{\text{d} #1}{\text{d}#2}}
\newcommand{\dde}[2]{\frac{\text{d}^2 #1}{\text{d} #2 ^2}}
\newcommand{\be}{\begin{equation}}
\newcommand{\ee}{\end{equation}}

\newcommand{\ppp}[3]{\frac{\partial^{2} #1}{\partial #2 \partial #3}}

\newcommand{\mS}{\mathcal{S}}
\newcommand{\mT}{\mathcal{T}}
\newcommand{\mM}{\mathcal{M}}
\title{The expressions for the $2^{nd}$-order mixed partial derivatives of Slater-Koster matrix elements at spherical coordinate singularities}
\author{Matthias Meister\footnote{\texttt{email:~M.Meister@qub.ac.uk}}\\
Atomistic Simulation Centre, Department of Physics and Astronomy\\ Queen's University Belfast,
Belfast BT7 1NN, United Kingdom, EU}
\begin{document}
\maketitle
\begin{abstract}
In a recent publication it has been shown how to generate derivatives with respect to atom coordinates of Slater-Koster 
matrix elements for the tight binding (TB) modelling of a system. For the special case of a mixed second partial derivative at 
coordinate singularities only the results were stated in that publication. In this work, the derivation of these results is given in
detail. Though it may seem rather `technical' and only applicable to a very special case, atomic configurations where the connecting 
vector between the two atoms involved in a two-centre matrix element is aligned along the $z$-axis (in the usual approach) require 
results for precisely this case. The expressions derived in this work have been implemented in the DINAMO code.   
\end{abstract}
\section{Slater-Koster matrix elements}
In \cite{Pod} it was shown how the Slater-Koster matrix element between orbitals localised at two different atoms could be systematically 
expressed in terms of the distance $R$ between the atoms, and the Euler angles $\alpha$ and $\beta$ describing the orientation of the 
connecting vector $\vec{R}=(X,Y,Z)=R(\cos\alpha\sin\beta,\sin\alpha\sin\beta,\cos\beta)$ in space and parametrising the rotations of the
original coordinate system into a new one necessary for the evaluation of the matrix element.
The general expression for a matrix element $\bra{l_{1}m_{1}}H\ket{l_{2}m_{2}}$, 
according to \cite{Pod}, with $l_{<}=\min(l_{1},l_{2})$, is
\be
\label{ME}
\begin{split}
\bra{l_1m_1}H\ket{l_2m_2}&(\alpha,\beta,R)=\sum_{m^{\prime}=1}^{l_<}
\biggl[S_{m_1m^{\prime}}^{l_1}(\alpha,\beta)S_{m_2m^{\prime}}^{l_2}(\alpha,\beta)
+T_{m_1m^{\prime}}^{l_1}(\alpha,\beta)T_{m_2m^{\prime}}^{l_2}(\alpha,\beta)\biggr]\\
&\qquad\times(l_1l_2|m^{\prime}|)(R)+
2A_{m_1}(\alpha)A_{m_2}(\alpha)d_{|m_1|0}^{l_1}(\beta)d_{|m_2|0}^{l_2}(\beta)(l_1l_20)(R)
\end{split}
\ee
Here $\ket{l_{i}m_{i}}$ denotes a state whose wavefunction has an angular dependence characterised by a real spherical harmonic
$\overline{Y}_{l_{i}m_{i}}$ on atoms $i=1,2$, respectively. The real spherical harmonics are defined as
\be
\overline{Y}_{lm}=\delta_{m0}Y_{l0}+(1-\delta_{m0})\sqrt{2}(-1)^m\big[\tau(m)\text{Re}Y_{l|m|}+\tau(-m)\text{Im}Y_{l|m|}\big]
\ee
with $\tau(m)=1$ if $m\geq 0$ and $\tau(m)=0$ if $m<0$. The $Y_{lm}$ are the ordinary complex valued spherical harmonics, with
the phase convention $Y_{lm}^{*}(\theta,\varphi)=(-1)^mY_{l-m}(\theta,\varphi)$.

$H$ in \gref{ME} is an operator that is cylindrically symmetric about all possible orientations of the connecting vector
$\vec{R}$; it can be the Hamiltonian, but also the identity, in which case the matrix elements would be overlap integrals between localised
states.
The other quantities appearing are
\be
A_m(\alpha):=\left\{ \begin{array}{ll}
(-1)^m\bigl[\tau(m)\cos(|m|\alpha)+\tau(-m)\sin(|m|\alpha)\bigr] & \textrm{if \(m\neq0\)}\\
\frac{1}{\sqrt{2}} & \textrm{if } m=0
\end{array}\right.
\ee
\be
B_m(\alpha):=\left\{\begin{array}{ll}(-1)^m\bigl[
  \tau(-m)\cos(|m|\alpha)-\tau(m)\sin(|m|\alpha)\bigr]&\textrm{if \(m\neq0\)}\\
0 & \textrm{if } m=0
\end{array}\right.
\ee
\be
\label{Sdef}
S_{mm^{\prime}}^l:=A_m\big[(-1)^{m^{\prime}} d_{|m|m^{\prime}}^l+d_{|m|-m^{\prime}}^l\big]
\ee
\be
\label{Tdef}
T_{mm^{\prime}}^l:=B_m\big[(-1)^{m^{\prime}}d_{|m|m^{\prime}}^l
-d_{|m|-m^{\prime}}^l\big]
\ee
and the Wigner $d$-function \cite{Var}
\be
\label{dwigner}
\begin{split}
d_{mm^{\prime}}^l(\beta)&=2^{-l}(-1)^{l-m^{\prime}}
\bigl[(l+m)!(l-m)!(l+m^{\prime})!(l-m^{\prime})!\bigr]^{\frac{1}{2}}\\
&\qquad\times\sum_{k=k_>}^{k_<}\frac{(-1)^k(1-\cos\beta)^{l-k-\frac{m+m^{\prime}}{2}}
(1+\cos\beta)^{k+\frac{m+m^{\prime}}{2}}}{k!(l-m-k)!(l-m^{\prime}-k)!(m+m^{\prime}+k)!}
\end{split}
\ee
with $k_<:=\min (l-m,l-m^{\prime}),\; k_>:=\max (0,-m-m^{\prime})$.
The $(l_{1}l_{2}|m'|)$ are the fundamental matrix elements, which depend only on the distance $R$ between the two atoms considered; they
give the matrix elements $\bra{l_{1}m'}H\ket{l_{2}m'}$ when the connecting vector is aligned with the $z$-axis, $\vec{R}=R\vec{e}_{z}$.
\\
In \cite{Elena} the approach of \cite{Pod} has been extended to $1^{st}$ and $2^{nd}$ order derivatives of the matrix element with respect 
to the Cartesian coordinates $X,Y,Z$ of the connecting vector $\vec{R}$. The approach detailed in \cite{Elena} can be applied to obtain
all first and second partial derivatives for all $\alpha, \beta, R$, with the exception of $\partial^{2} / \partial X\partial Y$ at 
 the poles $\beta=0,\pi$.  
This is due to a coordinate singularity and for the specific derivative mentioned no way around this problem was found 
within the approach of \cite{Elena}; a recourse to a Cartesian description was necessary. The calculations, however, were not detailed 
in that publication. This will be done in what follows.

The general expression for a Slater-Koster matrix element is a linear combination of $R$-dependent functions, 
the $(l_{1}l_{2}|m|)$, where the coefficients are combinations of integer powers of the direction 
cosines $X/R, Y/R, Z/R$ of the connecting vector. Therefore, the general form of a matrix element can 
be expressed as
\be
\label{SKM}
\mathcal{M}:=\bk{l_{1}m_{1}}{l_{2}m_{2}}=\sum_{ijk\geq 0}X^{i}Y^{j}Z^{k}f_{ijk}(R)
\ee
where $f_{ijk}$ is a combination of fundamental matrix elements $(l_{1}l_{2}|m|)$ with constant coefficients,
divided by $R^{i+j+k}$.
At the poles $Z=\pm R$, $X=Y=0$ only certain terms from the sum \gref{SKM} will remain in 
$\partial_{X}\partial_{Y}\mathcal{M}$. These are precisely those terms where $X^{i}Y^{j}Z^{k}=XYZ^{k}$, as we  
now show:
\begin{displaymath}
\partial_{X}\mathcal{M}=\sum_{ijk\geq 0}\left[iX^{i-1}Y^{j}Z^{k}f_{ijk}(R)+
X^{i+1}Y^{j}Z^{k}\frac{1}{R}\de{}{R}f_{ijk}(R)\right]
\end{displaymath}
Note that this is also valid if $i=0$.
\begin{multline*}
\partial_{Y}\partial_{X}\mathcal{M}=\sum_{ijk\geq 0}\left[ijX^{i-1}Y^{j-1}Z^{k}f_{ijk}(R)+
iX^{i-1}Y^{j+1}Z^{k}\frac{1}{R}\de{}{R}f_{ijk}(R)+\right.\\\left.
jX^{i+1}Y^{j-1}Z^{k}\frac{1}{R}\de{}{R}f_{ijk}(R)+
X^{i+1}Y^{j+1}Z^{k}\left(\frac{1}{R^{2}}\dde{}{R}f_{ijk}(R)-\frac{1}{R^{3}}\de{}{R}f_{ijk}(R)
\right)\right]
\end{multline*}
This is also valid if $j=0$ and/or $i=0$. As $i,j\geq 0$ it follows that for $X=Y=0$, $Z=\pm R$ the 2nd, 
3rd and 4th term vanish. The first term is non-zero in this case only if $i=j=1$. This is the proposition.
\section{Intermediate calculations}
It is convenient to rewrite the general expression \gref{ME} as
\begin{multline}
\label{ME2}
\mM=\bra{l_{1}m_{1}}H\ket{l_{2}m_{2}}=2A_{m_{1}}A_{m_{2}}d^{l_{1}}_{|m_{1}|0}d^{l_{2}}_{|m_{2}|0}(l_{1}l_{2}0)+
\sum_{m^{\prime}=1}^{l_{<}}\left[A_{m_{1}}A_{m_{2}}\mS^{l_{1}}_{m_{1}m^{\prime}}\mS^{l_{2}}_{m_{2}m^{\prime}}
+\right.\\\left.
B_{m_{1}}B_{m_{2}}\mT^{l_{1}}_{m_{1}m^{\prime}}\mT^{l_{2}}_{m_{2}m^{\prime}}\right]
(l_{1}l_{2}|m^{\prime}|)
\end{multline}
with $S^{l}_{mm^{\prime}}=A_{m}\mS^{l}_{mm^{\prime}}$ and $T^{l}_{mm^{\prime}}=B_{m}\mT^{l}_{mm^{\prime}}$.
We have for $|m_{1}|>0$ and $|m_{2}|>0$:
\begin{multline}
\label{Aprod}
A_{m_{1}}A_{m_{2}}=(-1)^{m_{1}+m_{2}}\left[\tau(m_{1})\cos(|m_{1}|\alpha)+\tau(-m_{1})\sin(|m_{1}|\alpha)
\right]\times\\\left[\tau(m_{2})\cos(|m_{2}|\alpha)+\tau(-m_{2})\sin(|m_{2}|\alpha)\right]=\\
(-1)^{m_{1}+m_{2}}\left[\tau(m_{1})\tau(m_{2})\cos(|m_{1}|\alpha)\cos(|m_{2}|\alpha)+
\tau(-m_{1})\tau(-m_{2})\sin(|m_{1}|\alpha)\sin(|m_{2}|\alpha)+\right.\\\left.
\tau(-m_{1})\tau(m_{2})\sin(|m_{1}|\alpha)\cos(|m_{2}|\alpha)+
\tau(m_{1})\tau(-m_{2})\cos(|m_{1}|\alpha)\sin(|m_{2}|\alpha)\right]
\end{multline}
and
\begin{multline}
\label{Bprod}
B_{m_{1}}B_{m_{2}}=(-1)^{m_{1}+m_{2}}\left[\tau(-m_{1})\cos(|m_{1}|\alpha)-\tau(m_{1})\sin(|m_{1}|\alpha)
\right]\times\\\left[\tau(-m_{2})\cos(|m_{2}|\alpha)-\tau(m_{2})\sin(|m_{2}|\alpha)\right]=\\
(-1)^{m_{1}+m_{2}}\left[\tau(-m_{1})\tau(-m_{2})\cos(|m_{1}|\alpha)\cos(|m_{2}|\alpha)+
\tau(m_{1})\tau(m_{2})\sin(|m_{1}|\alpha)\sin(|m_{2}|\alpha)-\right.\\\left.
\tau(-m_{1})\tau(m_{2})\cos(|m_{1}|\alpha)\sin(|m_{2}|\alpha)-
\tau(m_{1})\tau(-m_{2})\sin(|m_{1}|\alpha)\cos(|m_{2}|\alpha)\right]
\end{multline}
For $m\neq 0$ we obtain
\be
\label{NullProd}
A_{0}A_{m}=\frac{(-1)^{m}}{\sqrt{2}}\left[\tau(m)\cos(|m|\alpha)+\tau(-m)\sin(|m|\alpha)\right]
\ee
and $B_{0}B_{m}=0$.

A prefactor $XYZ^{k}$ in the terms of \gref{SKM} requires an $\alpha$-dependence 
that takes the form $\sin(2\alpha)=2\sin\alpha\cos\alpha$. If we convert the products of two functions $\cos$, $\sin$ 
in \gref{Aprod},\gref{Bprod} into sums of two such functions according to the usual trigonometric relations, we obtain $\sin$ functions
only for products of the form $\sin\cos$, according two
\begin{equation}
\label{trigo}
\begin{split}
\sin(|m_{1}|\alpha)\cos(|m_{2}|\alpha)&=\frac{1}{2}\sin[(|m_{2}|+|m_{1}|)\alpha]-\frac{1}{2}\sin[(|m_{2}|-|m_{1}|)\alpha]\\
\sin(|m_{2}|\alpha)\cos(|m_{1}|\alpha)&=\frac{1}{2}\sin[(|m_{2}|+|m_{1}|)\alpha]+\frac{1}{2}\sin[(|m_{2}|-|m_{1}|)\alpha]
\end{split}
\end{equation}
Henceforth we shall take $|m_{2}|\geq|m_{1}|$; this is no loss of generality, due to the symmetry properties of the inner product.
From \gref{trigo} and \gref{NullProd} we see that we can get the desired $\sin(2\alpha)$ for 
\begin{tabular}{ll}%
Case A & $|m_{1}|=|m_{2}|=1$\\
Case B.1 & $|m_{2}|-|m_{1}|=2,\;m_{1}=0$\\
Case B.2 & $|m_{2}|-|m_{1}|=2,\;|m_{1}|>0$
\end{tabular}\\
The distinction within case B is warranted for by the distinction in the coefficients $A_{m}, B_{m}$ for $m=0$ and $m>0$.
The $\tau$ functions in \gref{Aprod},\gref{Bprod},\gref{NullProd} further restrict the possible choices of $m_{1}$ and $m_{2}$.
In case A, we must have $m_{1}m_{2}=-1$, and we can choose, without loss of generality and without violating previous conventions
$m_{1}=1,m_{2}=-1$.
For B.1 we must have $m_{2}=-2$, and for B.2 $m_{1}m_{2}<0$. 
The matrix elements that can give contributions to the mixed partial derivative at the poles thus are:
$\bra{l_{1},1}H\ket{l_{2},-1}$, $\bra{l_{1},0}H\ket{l_{2},-2}$, and $\bra{l_{1},m_{1}}H\ket{l_{2},m_{2}}$ with $|m_{1}|>0$, 
$|m_{2}|=|m_{1}|+2$, and $m_{1}m_{2}<0$.

In order to identify these contributions we note that $XY=R^{2}(\sin\beta)^{2}\cos\alpha\sin\alpha$. 
We can find the required derivatives if we extract from \gref{ME2} the factor $(\sin\beta)^{2}\cos\alpha\sin\alpha$ where possible, 
evaluate the `remainder' at the poles and divide this result by $R^{2}$. This is equivalent to dividing by $XY$, and for a dependence
$XYZ^{k}$, the only one giving rise to nonvanishing contributions,  is also the same as taking the derivative with respect to $X$ and $Y$
at the poles.  
The $\beta$-dependence in the matrix elements stems from products of Wigner $d$-functions,
as can be seen from \gref{ME},\gref{Sdef},\gref{Tdef}.
The following relations will be needed:
\be
\label{Symrel}
\begin{split}
d^{l}_{mm'}(\beta)&=(-1)^{m-m'}d^{l}_{m'm}(\beta)\\
d^{l}_{mm'}(\pi-\beta)&= (-1)^{l+m}d^{l}_{m-m'}(\beta)
\end{split}
\ee

The $d$-functions $d^{l}_{mm'}$ occuring in the matrix element all have $m\geq 0$, and with \gref{Symrel} we can reformulate 
all expressions such that $m'\geq 0$ and $m\leq m'$. We therefore choose to rewrite \gref{dwigner} for $m,n\geq 0$ as
\begin{multline}
\label{dwignew}
d^{l}_{m(m+n)}(\vartheta)=2^{-l}(-1)^{l-m-n}\big[(l+m)!(l-m)!(l+m+n)!(l-m-n)!\big]^{\frac{1}{2}}\\%
\times\sum_{k=0}^{l-m-n}\frac{%
(-1)^{k}(1-\cos\vartheta)^{l-k-m-\frac{n}{2}}(1+\cos\vartheta)^{k+m+\frac{n}{2}}}%
{k!(l-m-k)!(l-m-n-k)!(2m+n+k)!}=\\
2^{-l}(-1)^{l-m-n}\big[(l+m)!(l-m)!(l+m+n)!(l-m-n)!\big]^{\frac{1}{2}}\\%
\times(1-\cos\vartheta)^{\frac{n}{2}}(1+\cos\vartheta)^{\frac{n}{2}}\sum_{k=0}^{l-m-n}\frac{%
(-1)^{k}(1-\cos\vartheta)^{l-m-n-k}(1+\cos\vartheta)^{k+m}}%
{k!(l-m-k)!(l-m-n-k)!(2m+n+k)!}=\\
(\sin\vartheta)^{n}(1+\cos\vartheta)^{m}2^{-l}(-1)^{l-m-n}\big[(l+m)!(l-m)!(l+m+n)!(l-m-n)!\big]^{\frac{1}{2}}\\%
\times\sum_{k=0}^{l-m-n}\frac{%
(-1)^{k}(1-\cos\vartheta)^{l-m-n-k}(1+\cos\vartheta)^{k}}%
{k!(l-m-k)!(l-m-n-k)!(2m+n+k)!}
\end{multline}
Only the behaviour of these functions at $\vartheta=0,\pi$ is important.
At $\vartheta=0$ only the term with $k=l-m-n$ contributes, and gives (where we don't evaluate $\sin\vartheta$ 
and $\cos\vartheta$ outside the sum)
\be
\label{angle0}
(\sin\vartheta)^{n}(1+\cos\vartheta)^{m}2^{-(m+n)}\sqrt{\frac{%
(l+m+n)!(l-m)!}{(l+m)!(l-m-n)!}}\frac{1}{n!}
\ee 
At $\vartheta=\pi$ the only contribution comes from $k=0$ and reads
\be
\label{anglepi}
(\sin\vartheta)^{n}(1+\cos\vartheta)^{m}2^{-(m+n)}(-1)^{l-m-n}\sqrt{\frac{%
(l+m)!(l+m+n)!}{(l-m)!(l-m-n)!}}\frac{1}{(2m+n)!}
\ee

We now look separately at the cases A, B.1, B.2, extracting $(\sin\beta)^{2}\cos\alpha\sin\alpha$.
\subsection{Case A : $m_{1}=1$, $m_{2}=-1$}
The relevant terms are
\be
\label{l1-c1}
2A_{1}A_{-1}d^{l_{1}}_{10}d^{l_{2}}_{10}(l_{1}l_{2}0)=
2\cos\alpha\sin\alpha d^{l_{1}}_{10}d^{l_{2}}_{10}(l_{1}l_{2}0)=2\cos\alpha\sin\alpha d^{l_{1}}_{01}d^{l_{2}}_{01}(l_{1}l_{2}0)
\ee
and
\begin{multline}
\label{l1-c2}
\sin\alpha\cos\alpha \sum_{m^{\prime}=1}^{l_{<}}\left[\mS^{l_{1}}_{1m^{\prime}}\mS^{l_{2}}_{-1m^{\prime}}-
\mT^{l_{1}}_{1m^{\prime}}\mT^{l_{2}}_{-1m^{\prime}}\right](l_{1}l_{2}|m^{\prime}|)=\\2\sin\alpha\cos\alpha
\sum_{m'=1}^{l_{<}}(-1)^{m'}\left[d^{l_{1}}_{1m^{\prime}}(\beta)d^{l_{2}}_{1-m^{\prime}}(\beta)+d^{l_{1}}_{1-m^{\prime}}(\beta)
d^{l_{2}}_{1m^{\prime}}(\beta)\right](l_{1}l_{2}|m^{\prime}|)=\\
2\sin\alpha\cos\alpha\sum_{m'=1}^{l_{<}}\left[(-1)^{l_{2}+m'+1}d^{l_{1}}_{1m'}(\beta)d^{l_{2}}_{1m'}(\pi-\beta)+
(-1)^{l_{1}+m'+1}d^{l_{1}}_{1m'}(\pi-\beta)d^{l_{2}}_{1m'}(\beta)\right](l_{1}l_{2}|m^{\prime}|)
\end{multline}
In the last equation the following relations have been used
\be
\label{useful}
\begin{split}
\mS^{l_{1}}_{1m^{\prime}}\mS^{l_{2}}_{-1m^{\prime}}&=d^{l_{1}}_{1m^{\prime}}d^{l_{2}}_{1m^{\prime}}+
d^{l_{1}}_{1-m^{\prime}}d^{l_{2}}_{1-m^{\prime}}+(-1)^{m^{\prime}}d^{l_{1}}_{1m^{\prime}}
d^{l_{2}}_{1-m^{\prime}}+(-1)^{m^{\prime}}d^{l_{1}}_{1-m^{\prime}}d^{l_{2}}_{1m^{\prime}}\\
\mT^{l_{1}}_{1m^{\prime}}\mT^{l_{2}}_{-1m^{\prime}}&=d^{l_{1}}_{1m^{\prime}}d^{l_{2}}_{1m^{\prime}}+
d^{l_{1}}_{1-m^{\prime}}d^{l_{2}}_{1-m^{\prime}}-(-1)^{m^{\prime}}d^{l_{1}}_{1m^{\prime}}
d^{l_{2}}_{1-m^{\prime}}-(-1)^{m^{\prime}}d^{l_{1}}_{1-m^{\prime}}d^{l_{2}}_{1m^{\prime}}
\end{split}
\ee
and \gref{Symrel} has been applied.

Using \gref{dwignew} we get at $\beta=0$ for \gref{l1-c1}
\be
\cos\alpha\sin\alpha(\sin\beta)^{2}\frac{1}{2}\sqrt{(l_{1}+1)l_{1}(l_{2}+1)l_{2}}(l_{1}l_{2}0)
\ee
and $(-1)^{l_{1}+l_{2}}\times$ this result at $\beta=\pi$.

For \gref{l1-c2} we see from \gref{dwignew} that in the case $n>0$ the product $d^{l_{1}}_{1m'}(\beta)d^{l_{2}}_{1m'}(\pi-\beta)$
contains at least a factor $(\sin\beta)^4$. As we only extract $(\sin\beta)^{2}$, the remainder vanishes at $\beta=0$ and $\beta=\pi$.
So only the case $m'=1$ can produce a contributing term; it is
\be
\sin\alpha\cos\alpha(\sin\beta)^{2}\big(-\frac{1}{4}\big)[l_{2}(l_{2}+1)+l_{1}(l_{1}+1)](l_{1}l_{2}1)
\ee 
at $\beta=0$ and $(-1)^{l_{1}+l_{2}}\times$ this expression at $\beta=\pi$.
\subsection{Case B.1 : $m_{1}=0$, $m_{2}=-2$}
The relevant terms are
\be
\label{l2-c1}
 2A_{0}A_{-2}d^{l_{1}}_{00}d^{l_{2}}_{20}(l_{1}l_{2}0)=
2\sqrt{2}\sin\alpha\cos\alpha d^{l_{1}}_{00}d^{l_{2}}_{02}(l_{1}l_{2}0)
\ee
and if $l_{1}>0$ also
\begin{multline}
\label{l2-c2}
A_{0}A_{-2}\sum_{m^{\prime}=1}^{l_{<}}
\mS^{l_{1}}_{0m^{\prime}}\mS^{l_{2}}_{-2m^{\prime}}(l_{1}l_{2}|m^{\prime}|)=\sqrt{2}\sin\alpha\cos\alpha
\sum_{m^{\prime}=1}^{l_{<}}\mS^{l_{1}}_{0m^{\prime}}\mS^{l_{2}}_{-2m^{\prime}}(l_{1}l_{2}|m^{\prime}|)=\\
\sqrt{2}\sin\alpha\cos\alpha
\sum_{m^{\prime}=1}^{l_{<}}\big[d^{l_{1}}_{0m^{\prime}}(\beta)d^{l_{2}}_{2m^{\prime}}(\beta)+
d^{l_{1}}_{0-m^{\prime}}(\beta)d^{l_{2}}_{2-m^{\prime}}(\beta)+\\(-1)^{m^{\prime}}d^{l_{1}}_{0m^{\prime}}(\beta)
d^{l_{2}}_{2-m^{\prime}}(\beta)
+(-1)^{m^{\prime}}d^{l_{1}}_{0-m^{\prime}}(\beta)d^{l_{2}}_{2m^{\prime}}(\beta)\big](l_{1}l_{2}|m^{\prime}|)=\\
\sqrt{2}\sin\alpha\cos\alpha\sum_{m^{\prime}=1}^{l_{<}}
\big[d^{l_{1}}_{0m^{\prime}}(\beta)d^{l_{2}}_{2m^{\prime}}(\beta)+
(-1)^{l_{1}+l_{2}}d^{l_{1}}_{0m'}(\pi-\beta)d^{l_{2}}_{2m'}(\pi-\beta)+\\
(-1)^{m'+l_{2}}d^{l_{1}}_{0m'}(\beta)d^{l_{2}}_{2m'}(\pi-\beta)+
(-1)^{m'+l_{1}}d^{l_{1}}_{0m'}(\pi-\beta)d^{l_{2}}_{2m'}(\beta)\big](l_{1}l_{2}|m^{\prime}|)
\end{multline}

Equation \gref{l2-c1} gives at $\beta=0$
\be
\sin\alpha\cos\alpha(\sin\beta)^{2}\frac{1}{2\sqrt{2}}\sqrt{(l_{2}+2)(l_{2}+1)l_{2}(l_{2}-1)}(l_{1}l_{2}0)
\ee
and at $\beta=\pi$ we obtain this expression $\times (-1)^{l_{1}+l_{2}}$.

Turning to \gref{l2-c2} we first note from \gref{dwignew} that for $m'\geq 3$ the products $d^{l_{1}}_{0m'}d^{l_{2}}_{2m'}$ contain 
at least a factor $(\sin\beta)^{4}$ and thus after extraction of $(\sin\beta)^{2}$ still vanish at $\beta=0,\pi$. This leaves us
with $m'=1,2$. We find:
\be
\label{ZW1}
d^{l_{1}}_{01}(\beta)d^{l_{2}}_{21}(\beta)=-d^{l_{1}}_{01}(\beta)d^{l_{2}}_{12}(\beta)=-(\sin\beta)^{2}
\frac{1}{4}\sqrt{(l_{1}+1)l_{1}(l_{2}+2)(l_{2}-1)}
\ee 
at $\beta=0$, whereas this product vanishes at $\beta=\pi$ after extraction of $(\sin\beta)^{2}$ due to a factor $(1+\cos\beta)$.
$d^{l_{1}}_{01}(\pi-\beta)d^{l_{2}}_{21}(\pi-\beta)$ vanishes at $\beta=0$ due to a factor $(1-\cos\beta)$, and at $\beta=\pi$ results in
\gref{ZW1}.

$d^{l_{1}}_{01}(\beta)d^{l_{2}}_{21}(\pi-\beta)$ vanishes at $\beta=0$ and at $\beta=\pi$ gives
\be
\label{ZW2}
d^{l_{1}}_{01}(\beta)d^{l_{2}}_{21}(\pi-\beta)=-d^{l_{1}}_{01}d^{l_{2}}_{12}(\pi-\beta)=(-1)^{l_{1}}(\sin\beta)^{2}\frac{1}{4}
\sqrt{(l_{1}+1)l_{1}(l_{2}+2)(l_{2}-1)}
\ee
$d^{l_{1}}_{01}(\pi-\beta)d^{l_{2}}_{21}(\beta)$ yields \gref{ZW2} at $\beta=0$ and vanishes at $\beta=\pi$.

So from $m'=1$ we get a contribution  which at $\beta=0$ is
\be
\sin\alpha\cos\alpha(\sin\beta)^{2}\frac{-1}{\sqrt{2}}\sqrt{(l_{1}+1)l_{1}(l_{2}+2)(l_{2}-1)}(l_{1}l_{2}1)
\ee
and at $\beta=\pi$ just $(-1)^{l_{1}+l_{2}}\times$ this result.

At $\beta=0$
\be
d^{l_{1}}_{02}(\beta)d^{l_{2}}_{22}(\beta)=\frac{1}{8}(\sin\beta)^{2}\sqrt{(l_{1}+2)(l_{1}+1)l_{1}(l_{1}-1)}
\ee
whereas this product does not contribute at $\beta=\pi$; there $d^{l_{1}}_{02}(\pi-\beta)d^{l_{2}}_{22}(\pi-\beta)$ gives the
above expression and in turn vanishes at $\beta=0$. 
At $\beta=0$ $d^{l_{1}}_{02}(\beta)d^{l_{2}}_{22}(\pi-\beta)$ vanishes and at $\beta=\pi$ yields
\be
(-1)^{l_{1}}\frac{1}{8}(\sin\beta)^{2}\sqrt{(l_{1}+2)(l_{1}+1)l_{1}(l_{1}-1)}
\ee
Vice versa $d^{l_{1}}_{02}(\pi-\beta)d^{l_{2}}_{22}(\beta)$ gives the above expression at $\beta=0$ and vanishes at $\beta=\pi$.
Therefore the $m'=2$ contribution is
\be
\sin\alpha\cos\alpha(\sin\beta)^{2}\frac{1}{2\sqrt{2}}\sqrt{(l_{1}+2)(l_{1}+1)l_{1}(l_{1}-1)}(l_{1}l_{2}2)
\ee
and at $\beta=\pi$ just $(-1)^{l_{1}+l_{2}}\times$ this result.
\subsection{Case B.2 : $|m_{1}|>0$, $|m_{2}|=|m_{1}|+2$}
There will be no contribution from the $(l_{1}l_{2}0)$ term in \gref{ME2}, because 
\be
d^{l_{1}}_{0|m_{1}|}d^{l_{2}}_{0|m_{2}|}\propto (\sin\beta)^{|m_{1}|+|m_{2}|}=(\sin\beta)^{2|m_{1}|+2}
\ee
and, because $|m_{1}|>0$, the remainder after extraction of $(\sin\beta)^{2}$ vanishes at $\beta=0,\pi$.
In the sum over $(l_{1}l_{2}|m'|)$ for $m'>0$ only those parts of the prefactors will contribute to the derivative we seek 
that depend on $\alpha$ like $\cos\alpha\sin\alpha$. If we drop all other terms, this `reduced' sum $\tilde{\mathcal{M}}$ 
reads in the case $m_{1}>0$, $m_{2}<0$
\be
\label{RS}
\tilde{\mathcal{M}}=(-1)^{m_{1}+m_{2}}\sin\alpha\cos\alpha\sum_{m'=1}^{l_{<}}\big[%
\mS^{l_{1}}_{m_{1}m'}\mS^{l_{2}}_{m_{2}m'}+\mT^{l_{1}}_{m_{1}m'}\mT^{l_{2}}_{m_{2}m'}\big](l_{1}l_{2}|m'|)
\ee
and in the case $m_{1}<0$, $m_{2}>0$ it is just the negative of \gref{RS}. We will be using $m_{1}>0$, $m_{2}<0$ until stated otherwise.
With \gref{Symrel} we can write
\begin{multline}
\mS^{l_{1}}_{m_{1}m'}(\beta)\mS^{l_{2}}_{m_{2}m'}(\beta)=
d^{l_{1}}_{|m_{1}|m'}(\beta)d^{l_{2}}_{|m_{2}|m'}(\beta)+\\(-1)^{l_{1}+l_{2}+|m_{1}|+|m_{2}|}
d^{l_{1}}_{|m_{1}|m'}(\pi-\beta)d^{l_{2}}_{|m_{2}|m'}(\pi-\beta)+\\ 
(-1)^{l_{2}+|m_{2}|+m'}d^{l_{1}}_{|m_{1}|m'}(\beta)d^{l_{2}}_{|m_{2}|m'}(\pi-\beta)+\\
(-1)^{l_{1}+|m_{1}|+m'}d^{l_{1}}_{|m_{1}|m'}(\pi-\beta)d^{l_{2}}_{|m_{2}|m'}(\beta)
\end{multline}
\begin{multline}
\mT^{l_{1}}_{m_{1}m'}(\beta)\mT^{l_{2}}_{m_{2}m'}(\beta)=
d^{l_{1}}_{|m_{1}|m'}(\beta)d^{l_{2}}_{|m_{2}|m'}(\beta)+\\(-1)^{l_{1}+l_{2}+|m_{1}|+|m_{2}|}
d^{l_{1}}_{|m_{1}|m'}(\pi-\beta)d^{l_{2}}_{|m_{2}|m'}(\pi-\beta)-\\ 
(-1)^{l_{2}+|m_{2}|+m'}d^{l_{1}}_{|m_{1}|m'}(\beta)d^{l_{2}}_{|m_{2}|m'}(\pi-\beta)-\\
(-1)^{l_{1}+|m_{1}|+m'}d^{l_{1}}_{|m_{1}|m'}(\pi-\beta)d^{l_{2}}_{|m_{2}|m'}(\beta)
\end{multline}
Therefore
\begin{multline}
\label{Therefore}
\mS^{l_{1}}_{m_{1}m'}\mS^{l_{2}}_{m_{2}m'}+\mT^{l_{1}}_{m_{1}m'}\mT^{l_{2}}_{m_{2}m'}=\\
2d^{l_{1}}_{|m_{1}|m'}(\beta)d^{l_{2}}_{|m_{2}|m'}(\beta)+2(-1)^{l_{1}+l_{2}}
d^{l_{1}}_{|m_{1}|m'}(\pi-\beta)d^{l_{2}}_{|m_{2}|m'}(\pi-\beta)
\end{multline}
Note that $(-1)^{m_{1}+m_{2}}=(-1)^{|m_{1}|+|m_{2}|}=(-1)^{|m_{2}|-|m_{1}|}=1$ in the present case.
The second term in \gref{Therefore} involves the same functions as the first, only evaluated at $\pi-\beta$ instead of $\beta$. 
In order to evaluate\footnote{Always explicitly keeping $(\sin\beta)^{2}$.}
 the first term at $\beta=0$, we have to choose $\vartheta=\beta$ in \gref{angle0}; for $\beta=\pi$,  $\vartheta=\beta$ has to 
be referred to 
\gref{anglepi}, and we can see that this vanishes ($m\geq 1$).
For the second term at $\beta=0$, the required choice is $\vartheta=\pi-\beta$ in \gref{anglepi}. This expression vanishes, as 
$[1+\cos(\pi-\beta)]^m=[1-\cos(\beta)]^m\rightarrow 0$ as $\beta\rightarrow 0$.
At $\beta=\pi$ the second term is evaluated by plugging $\vartheta=\pi-\beta$ into \gref{angle0}. This evidently gives the same result as
putting $\beta=0$ in \gref{angle0}. Looking at \gref{Therefore} as a whole it follows that at $\beta=\pi$ it 
results in $(-1)^{l_{1}+l_{2}}\times$ its
value at $\beta=0$, the latter being determined entirely by the first term.

If $m'>|m_{2}|=|m_{1}|+2$ the product $d^{l_{1}}_{|m_{1}|m'}d^{l_{2}}_{|m_{2}|m'}$ contains at least a factor $(\sin\beta)^{4}$.
It therefore does not contribute. If $m'<|m_{1}|=|m_{2}|-2$ then
$d^{l_{1}}_{|m_{1}|m'}d^{l_{2}}_{|m_{2}|m'}=(-1)^{|m_{1}|+|m_{2}|}d^{l_{1}}_{m'|m_{1}|}d^{l_{2}}_{m'|m_{2}|}$ and this again contains at
least $(\sin\beta)^{4}$. So we have to consider explicitly the cases $m'=|m_{2}|$, $m'=|m_{1}|+1$, and $m'=|m_{1}|$, and only at $\beta=0$.

\underline{$m'=|m_{2}|$} :
\begin{multline}
\label{ZW3}
d^{l_{1}}_{|m_{1}||m_{2}|}(\beta)d^{l_{2}}_{|m_{2}||m_{2}|}(\beta)=\\
\frac{1}{8}(\sin\beta)^{2}\sqrt{(l_{1}+|m_{1}|+2)(l_{1}+|m_{1}|+1)(l_{1}-|m_{1}|)(l_{1}-|m_{1}|-1)}
\end{multline}

\underline{$m'=|m_{1}|+1=|m_{2}|-1$} :
\begin{multline}
\label{ZW4}
d^{l_{1}}_{|m_{1}|(|m_{1}|+1)}(\beta)d^{l_{2}}_{|m_{2}|(|m_{2}|-1)}(\beta)=
-d^{l_{1}}_{|m_{1}|(|m_{1}|+1)}(\beta)d^{l_{2}}_{(|m_{1}|+1)|m_{2}|}(\beta)=\\
-(\sin\beta)^{2}\frac{1}{4}\sqrt{(l_{1}+|m_{1}|+1)(l_{1}-|m_{1}|)(l_{2}+|m_{1}|+2)(l_{2}-|m_{1}|-1)}
\end{multline}

\underline{$m'=|m_{1}|$} :
\begin{multline}
\label{ZW5}
d^{l_{1}}_{|m_{1}||m_{1}|}(\beta)d^{l_{2}}_{|m_{2}||m_{1}|}(\beta)=d^{l_{1}}_{|m_{1}||m_{1}|}(\beta)d^{l_{2}}_{|m_{1}||m_{2}|}(\beta)=\\
\frac{1}{8}(\sin\beta)^{2}\sqrt{(l_{2}+|m_{1}|+2)(l_{2}+|m_{1}|+1)(l_{2}-|m_{1}|)(l_{2}-|m_{1}|-1)}
\end{multline}

%
%
%

\section{Evaluation of $\ppp{}{X}{Y}\bra{l_{1}m_{1}}H\ket{l_{2}m_{2}}$ at the poles}
With the results of the previous section we now compose expressions for the matrix elements at the poles.
As explained earlier, dropping $\cos\alpha\sin\alpha(\sin\beta)^{2}$ from these expressions and dividing the
remainder by $R^{2}$ gives the derivative we want. Thus we find at $\beta=0$:

Case A:
\begin{multline}
\label{Result1}
\partial_{X}\partial_{Y}\bra{l_{1},+1}H\ket{l_{2},-1}=\frac{1}{R^{2}}\left\{
\frac{1}{2}\sqrt{l_{1}(l_{1}+1)l_{2}(l_{2}+1)}\;(l_{1}l_{2}0)\right.\\\left.-\frac{1}{4}
[l_{1}(l_{1}+1)+l_{2}(l_{2}+1)](l_{1}l_{2}1)\right\}
\end{multline}

Case B.1:
\begin{multline}
\label{Result2}
\partial_{X}\partial_{Y}\bra{l_{1},0}H\ket{l_{2},-2}=
\frac{1}{2\sqrt{2}R^{2}}\sqrt{(l_{2}+2)(l_{2}+1)l_{2}(l_{2}-1)}\;(l_{1}l_{2}0)
\\
-(1-\delta_{l_{1}0})\frac{1}{\sqrt{2}R^{2}}
\sqrt{l_{1}(l_{1}+1)(l_{2}+2)(l_{2}-1)}\;
(l_{1}l_{2}1)\\
+(1-\delta_{l_{1}0})(1-\delta_{l_{1}1})\frac{1}{2\sqrt{2}R^{2}}
\sqrt{(l_{1}+2)(l_{1}+1)l_{1}(l_{1}-1)}\;(l_{1}l_{2}2)
\end{multline}

Case B.2 ($m_{1}>0$, $m_{2}<0$, otherwise multiply result below by $-1$) :
\begin{multline}
\label{Result3}
\partial_{X}\partial_{Y}\bra{l_{1},m_{1}}H\ket{l_{2},m_{2}}=\\\frac{1}{R^{2}}\Big[%
\frac{1}{4}\sqrt{(l_{2}+|m_{1}|+2)(l_{2}+|m_{1}|+1)(l_{2}-|m_{1}|)(l_{2}-|m_{1}|-1)}(l_{1}l_{2}|m_{1}|)\\
-(1-\delta_{l_{1}|m_{1}|})\frac{1}{2}\sqrt{(l_{1}+|m_{1}|+1)(l_{1}-|m_{1}|)(l_{2}+|m_{1}|+2)(l_{2}-|m_{1}|-1)}(l_{1}l_{2}(|m_{1}|+1))\\
+(1-\delta_{l_{1}|m_{1}|})(1-\delta_{l_{1}(|m_{1}|+1)})\frac{1}{4}
\sqrt{(l_{1}+|m_{1}|+2)(l_{1}+|m_{1}|+1)(l_{1}-|m_{1}|)(l_{1}-|m_{1}|-1)}(l_{1}l_{2}|m_{2}|)\Big]
\end{multline}
At $\beta=\pi$ we get $(-1)^{l_{1}+l_{2}}\times$ these values.

Expressions \gref{Result1} and \gref{Result2} were stated in \cite{Elena}. In this work their derivation 
has been demonstrated. Result \gref{Result3} was not mentioned in \cite{Elena}. 

\bigskip
{\bf Acknowledgements :} This work was supported by EPSRC under grant GR/S80165.


\begin{thebibliography}{00}
\bibitem{Pod} Podolskiy A V, Vogl P 2004 Phys. Rev. B {\bf 69} 233101
\bibitem{Elena} Elena A M, Meister M 2005  \texttt{cond-mat/0504719}
\bibitem{Var} Varshalovich et al. 1988 {\it Quantum Theory of Angular Momentum} (Singapore: World Scientific)
\end{thebibliography}
\end{document}